# Astro2020 Science White Paper

# Understanding the circumgalactic medium is critical for understanding galaxy evolution

**Thematic Areas:**
☐ Planetary Systems ☐ Star and Planet Formation
☐ Formation and Evolution of Compact Objects ☑ Cosmology and Fundamental Physics
☐ Stars and Stellar Evolution ☑ Resolved Stellar Populations and their Environments
☑ Galaxy Evolution ☐ Multi-Messenger Astronomy and Astrophysics

**Principal Author:**
Name: Molly S. Peeples
Institution: Space Telescope Science Institute / Johns Hopkins University
Email: molly@stsci.edu
Phone: 1 (410) 338-2451

**Co-authors:**
Peter Behroozi, University of Arizona, behroozi@email.arizona.edu
Rongmon Bordoloi, North Carolina State University, rbordol@ncsu.edu
Alyson Brooks, Rutgers, the State University of New Jersey, abrooks@physics.rutgers.edu
James S. Bullock, University of California, Irvine, bullock@uci.edu
Joseph N. Burchett, University of California, Santa Cruz, burchett@ucolick.org
Hsiao-Wen Chen, University of Chicago, hchen@oddjob.uchicago.edu
John Chisholm, University of California, Santa Cruz, jochisho@ucsc.edu
Charlotte Christensen, Grinnell College, christenc@grinnell.edu
Alison Coil, University of California, San Diego, acoil@ucsd.edu
Lauren Corlies, The Large Synoptic Survey Telescope, lcorlies@lsst.org
Aleksandar Diamond-Stanic, Bates College, adiamond@bates.edu
Megan Donahue, Michigan State University, donahue@pa.msu.edu
Claude-André Faucher-Giguère, Northwestern University, cgiguere@northwestern.edu
Henry Ferguson, Space Telescope Science Institute, ferguson@stsci.edu
Drummond Fielding, Flatiron Institute, dfielding@flatironinstitute.org
Andrew J. Fox, Space Telescope Science Institute, afox@stsci.edu
David M. French, Space Telescope Science Institute, dfrench@stsci.edu
Steven R. Furlanetto, University of California, Los Angeles, sfurlane@astro.ucla.edu
Mario Gennaro, Space Telescope Science Institute, gennaro@stsci.edu
Karoline M. Gilbert, Space Telescope Science Institute / Johns Hopkins University, kgilbert@stsci.edu
Erika Hamden, University of Arizona, hamden@email.arizona.edu
Nimish Hathi, Space Telescope Science Institute, nhathi@stsci.edu
Matthew Hayes, Stockholm University, matthew@astro.su.se

Alaina Henry, Space Telescope Science Institute, `ahenry@stsci.edu`
J. Christopher Howk, University of Notre Dame, `howk.1@nd.edu`
Cameron Hummels, California Institute of Technology, `chummels@caltech.edu`
Dušan Kereš, University of California, San Diego, `dukeres@ucsd.edu`
Evan Kirby, California Institute of Technology, `enk@astro.caltech.edu`
Anton M. Koekemoer, Space Telescope Science Institute, `koekemoer@stsci.edu`
Ting-Wen Lan, Kavli Institute for the Physics and Mathematics of the Universe, `tingwenlan@gmail.com`
Lauranne Lanz, Dartmouth College, `lauranne.lanz@dartmouth.edu`
David R. Law, Space Telescope Science Institute, `dlaw@stsci.edu`
Nicolas Lehner, University of Notre Dame, `nlehner@nd.edu`
Jennifer M. Lotz, Gemini Observatory, `jlotz@gemini.edu`
Crystal L. Martin, University of California, Santa Barbara, `cmartin@ucsb.edu`
Kristen McQuinn, Rutgers University, `kristen.mcquinn@rutgers.edu`
Matthew McQuinn, University of Washington, `mcquinn@uw.edu`
Ferah Munshi, University of Oklahoma, `ferahmunshi@gmail.com`
S. Peng Oh, University of California, Santa Barbara, `peng@physics.ucsb.edu`
John M. O'Meara, W. M. Keck Observatory, `jomeara@keck.hawaii.edu`
Brian W. O'Shea, Michigan State University, `oshea@msu.edu`
Camilla Pacifici, Space Telescope Science Institute, `cpacifici@stsci.edu`
J. E. G. Peek, Space Telescope Science Institute / Johns Hopkins University, `jegpeek@stsci.edu`
Marc Postman, Space Telescope Science Institute, `postman@stsci.edu`
Moire Prescott, New Mexico State University, `mkpresco@nmsu.edu`
Mary Putman, Columbia University, `mputman@astro.columbia.edu`
Eliot Quataert, University of California, Berkeley, `eliot@berkeley.edu`
Marc Rafelski, Space Telescope Science Institute / Johns Hopkins University, `mrafelski@stsci.edu`
Joseph Ribaudo, Utica College, `jsribaud@utica.edu`
Kate Rowlands, Johns Hopkins University, `katerowlands.astro@gmail.com`
Kate Rubin, San Diego State University, `krubin@sdsu.edu`
Brett Salmon, Space Telescope Science Institute, `bsalmon@stsci.edu`
Claudia Scarlata, University of Minnesota, `mscarlat@umn.edu`
Alice E. Shapley, University of California, Los Angeles, `aes@astro.ucla.edu`
Raymond Simons, Space Telescope Science Institute, `rsimons@stsci.edu`
Gregory F. Snyder, Space Telescope Science Institute, `gsnyder@stsci.edu`
Jonathan Stern, Northwestern University, `jonathan.stern@northwestern.edu`
Allison L. Strom, Carnegie Observatories, `astrom@carnegiescience.edu`
Erik Tollerud, Space Telescope Science Institute, `etollerud@stsci.edu`
Paul Torrey, University of Florida, `paul.torrey@ufl.edu`
Grant Tremblay, Harvard-Smithsonian Center for Astrophysics, `gtremblay@cfa.harvard.edu`
Todd M. Tripp, University of Massachusetts, Amherst, `tripp@astro.umass.edu`
Jason Tumlinson, Space Telescope Science Institute / Johns Hopkins University, `tumlinson@stsci.edu`

Sarah Tuttle, University of Washington / Apache Point Observatory, `tuttlese@uw.edu`
Frank C. van den Bosch, Yale University, `frank.vandenbosch@yale.edu`
G. Mark Voit, Michigan State University, `voit@pa.msu.edu`
Q. Daniel Wang, University of Massachusetts, Amherst, `wqd@umass.edu`
Jessica K. Werk, University of Washington, `jwerk@uw.edu`
Benjamin F. Williams, University of Washington, `benw1@uw.edu`
Dennis Zaritsky, University of Arizona, `dennis.zaritsky@gmail.com`
Yong Zheng, University of California, Berkeley, `yongzheng@berkeley.edu`

**Abstract:** Galaxies evolve under the influence of gas flows between their interstellar medium and their surrounding gaseous halos known as the circumgalactic medium (CGM). The CGM is a major reservoir of galactic baryons and metals, and plays a key role in the long cycles of accretion, feedback, and recycling of gas that drive star formation. In order to fully understand the physical processes at work *within* galaxies, it is therefore essential to have a firm understanding of the composition, structure, kinematics, thermodynamics, and evolution of the CGM. In this white paper we outline connections between the CGM and galactic star formation histories, internal kinematics, chemical evolution, quenching, satellite evolution, dark matter halo occupation, and the reionization of the larger-scale intergalactic medium in light of the advances that will be made on these topics in the 2020s. We argue that, in the next decade, fundamental progress on all of these major issues depends critically on improved empirical characterization and theoretical understanding of the CGM. In particular, we discuss how future advances in spatially-resolved CGM observations at high spectral resolution, broader characterization of the CGM across galaxy mass and redshift, and expected breakthroughs in cosmological hydrodynamic simulations will help resolve these major problems in galaxy evolution.

# Galaxy evolution in the 2020s

Throughout the late 2000s and 2010s, the study of galaxy evolution was revolutionized by advancements in integral-field spectroscopy that deconvolved physical processes on galactic scales. Where once galaxies had been studied only in terms of their globally-integrated properties, programs such as KMOS$^{3D}$ (Wisnioski et al., 2015) and MaNGA (Bundy et al., 2015) and instruments such as VLT/MUSE (Bacon et al., 2010) and Keck/MOSFIRE (McLean et al., 2012) provide spatially resolved spectroscopy for tens of thousands of galaxies across the universe. The 2020s will extend these successes with ground- and space-based integral field and multi-object spectrographs that will enable mapping of spatially resolved star formation and chemical histories to ever higher redshifts and fainter luminosities, coupled with a deeper understanding of the interstellar medium (ISM) and large-scale structure. In the nearby universe, large-area surveys combined with spatially-resolved spectroscopy (e.g., SDSS-V; Kollmeier et al., 2017) will connect global galaxy properties to star-forming regions and the gaseous ISM on scales from 0.1–1000 pc. Likewise, wide-field surveys (e.g., those with *WFIRST*; Akeson et al., 2019, and LSST; LSST Science Collaboration et al., 2017) will reveal stellar halo substructures and satellites of the Milky Way and nearby galaxies, further narrowing the gap between the fields of "resolved stellar populations" and "galaxy evolution." *JWST* (Kalirai, 2018) and the ELTs will in turn unveil the early universe, giving improved views of the internal kinematics and composition of galaxies at $z > 1$ and a clear picture of the first galaxies. Meanwhile, *JWST* and next-generation radio observatories (ngVLA, VLBI, SKA, ALMA) will map the dusty, atomic, and molecular ISM with unprecedented fidelity.

We posit here that though the 2020s will provide dramatic advancements in our empirical understanding of galaxy evolution, **attaining a complete physical picture of how galaxies evolve will be fundamentally limited without comparable improvements in our knowledge of the circumgalactic medium (CGM)**. The CGM hosts the large-scale gas flows driving galaxy evolution: it is the source of star-forming fuel, the immediate destination of galaxy-scale outflows, and the regulator between galaxies' dark matter halos and the intergalactic medium (IGM). Though the CGM is substantially less dense than the ISM, this vast reservoir harbors 3–5 times as many baryons *and* heavy elements as galaxies. We discuss here advances that will be made in seven galaxy evolution topics (see the Table) in the 2020s, open questions that will likely remain unanswered, and how the CGM is central to answering these questions. Topics #1–5 predominantly concern how galaxies themselves evolve, while #5–7 relate to using galaxies as cosmological probes. We conclude with a non-exhaustive list of improved CGM data and models that would directly address these questions.

**1. Morphologies and Internal Kinematics:** The increase of spatially-resolved spectroscopy continues to expand our understanding of the stellar and gaseous kinematics of galaxies at all epochs. The last decade has shown galactic disks are substantially more turbulent in the young universe than they are locally (e.g., Law et al., 2009; Wisnioski et al., 2015; Simons et al., 2017). *JWST* and ELTs will enable resolved velocity maps for hundreds of galaxies at $z > 2$, revealing if disordered motions at earlier epochs trace the cosmic star formation rate density or the cosmic merger rate. However, **how angular momentum is exchanged between the ISM and the CGM** will be critical to interpreting these observations. How gas is accreted onto the ISM (morphology, angular momentum, thermodynamics, etc.) affects where and when stars are formed (#2) and thus the dynamics of young stellar populations. Galaxies lose angular momentum via outflows, and whether, how, where, and when this material is recycled back onto the galaxy impact the full system dynamics. Mergers also drive kinematic evolution, and the CGM affects satellite galaxies (#5). Finally, star formation quenching (#4) is usually in conjunction with galaxies' kinematic transitions.

| | Galaxy Evolution Topic | 2020s Advances | Circumgalactic Relevance | Needed CGM Information |
|---|---|---|---|---|
| 1 | Morphologies and Internal Kinematics | Spatially resolved maps in ISM and stars to fainter sources, higher redshifts, smaller spatial scales | Angular momentum exchange; dynamical and thermal state of gas accreting onto ISM | Spatially resolved CGM kinematics with respect to galaxies' resolved kinematics |
| 2 | Star Formation Histories | Spatially resolved SFHs to smaller scales and higher redshifts; improved SED modeling | Fuel (or lack thereof) for star formation comes from the CGM | Accretion rates, morphologies, kinematics, and thermodynamics |
| 3 | Chemical Evolution | Spatially resolved ISM metallicity maps for more galaxies, earlier epochs; more resolved stellar populations in Milky Way and the local universe | Most heavy elements are in CGM. Accretion dilutes ISM; recycling & outflows redistribute metals. | Understand metallicity substructure & mixing; distribution & dynamics of metals |
| 4 | Quenching | Improved understanding of quenching *within* galaxies, environmental dependence, co-evolution with SMBHs | CGM drives or impacted by quenching event; source of any gas accreting onto galaxies | Thermodynamics of gas in the CGM of quenching and quenched galaxies |
| 5 | Satellite Evolution | Lower masses outside of Local Group; earlier epochs; more detailed SFHs | Satellites are traveling through central galaxies' CGM | Spatially and spectrally resolved maps to determine density and kinematic structure |
| 6 | The Dark Matter Halo Connection | Deep large-scale structure surveys in visible and near-infrared | CGM modulates how/if gas accreted onto halo reaches the ISM | CGM gas distribution for different kinds of galaxies at all epochs |
| 7 | Reionization | Map it; determine sources, morphology, redshift range, rate | Affects how much ionizing flux escaping galaxies reach the IGM | Small-scale structure and opacities to ionizing flux |

**2. Star Formation Histories:** The next decade will see a continued increase in the determination of spatially-resolved star formation histories (SFHs) via the increase of spatially-resolved spectrophotometric observations of distant galaxies and of resolved stellar populations of nearby galaxies. Combined with larger samples of galaxy gas masses, these data will help constrain the duty cycles of small- and large-scale stochasticity, yielding new insights into how galaxies evolve through the stellar mass–star formation rate plane and what sets the scatter in the so-called star forming main sequence. The 2010s saw major improvements in modeling full galaxy SEDs (Conroy, 2013) and in using these models to measure SFHs. Deep, resolved spectrophotometry across a much larger range of galaxy mass, morphology, and environments in the 2020s will further challenge and constrain SED models, but the relatively simple approaches of the last decade will no longer suffice for interpreting these results. A galaxy's SFH is a record of its gas accretion and merger history (see #5) coupled with its internal dynamics (#1) and outflows. **The rate, temperature, angular momentum, and location of accreting gas, in turn, is dictated by the physics of the CGM** (e.g., accretion dominated by collimated flows versus condensation). Therefore, to interpret galaxy SFHs, we must also understand how gas flows into galaxies from the CGM.

**3. Chemical Evolution:** Characterization of how, when, and where galaxies build up and redistribute their metals will improve in the 2020s thanks to several advances. Deep NIR spectra (from *JWST* and ELTs) will enable access to visible-wavelength emission lines at $z > 1$ for more and fainter galaxies than available currently (Sanders et al., 2015). The proliferation of spatially resolved emission line maps will help disentangle strong emission-line diagnostics and hence map metallicities with respect to galaxy kinematics (Poetrodjojo et al., 2018). Spectroscopic surveys of resolved stellar populations in the Local Volume will help place the evolution of individual elements in the broader context of detailed SFHs (#2). Yet **a full accounting of the heavy elements requires mapping the CGM: most metals are not in galaxies** (Peeples et al., 2014; Peek et al., 2015; Telford et al., 2018). Galaxies lose metals to the CGM via outflows (and, in smaller systems, gas stripping; #5); these metals may mix with the CGM before potentially recycling back into the ISM. Thus understanding the details of galactic chemical evolution requires understanding of small-scale circumgalactic gas physics. Conversely, metals are the dominant coolant in the CGM. Thermodynamics depends critically on the cooling rate (Smith et al., 2017): accretion physics—and so all the topics discussed here—is hence profoundly affected by CGM metals.

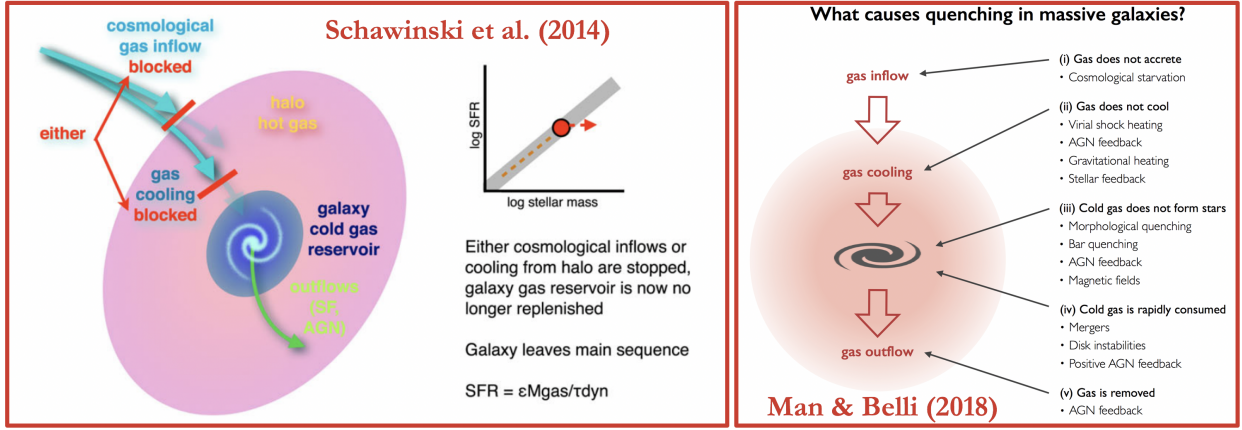

Figure 1: Galaxy quenching models have testable CGM predictions. Left: A proposed late-type galaxy quenching (Schawinski et al., 2014). Right: from the Man & Belli (2018) summary of the 2017 Lorentz Center workshop, "The Physics of Quenching Massive Galaxies at High Redshift".

**4. Quenching:** One of the most longstanding and perplexing questions in galaxy evolution is how, why, and when do (some) galaxies cease forming stars. Over the next decade, deeper and wider imaging surveys coupled with spatially-resolved spectroscopy will better map which galaxies quench when, and, combined with next-generation radio surveys, will better characterize the distribution and thermodynamics of gas in quenching and passive galaxies (Rowlands et al., 2015). Predominant theories postulate that some source of energy is required to prohibit interstellar gas from cooling, collapsing, and forming stars (Somerville & Davé, 2015): **essentially *all* theories for galaxy quenching either invoke changes to circumgalactic gas or have significant testable predictions for how the CGM changes throughout the quenching process** (Figure 1). A central tenant of many models is eliminating cool gas in the inner CGM, thereby stalling accretion of fresh star-forming fuel into the ISM. Yet *Hubble* observations reveal significant reservoirs of cool gas in the inner CGM of *passive* galaxies (Thom et al., 2012; Chen, 2017; Berg et al., 2018). Measuring the CGM cooling rate directly (via emission) can confirm the inhibition of cooling, perhaps due to non-thermal energy input that could be revealed by mapping circumgalactic dynamics.

**5. Satellites:** Satellite galaxy evolution is heavily influenced by the CGM of the host galaxy. It is not yet understood how satellite SFHs evolve once they "fall" into the central galaxy's dark matter halo (Wetzel et al., 2014), how and when they lose their ISM, whether they merge with the central or become part of the stellar halo, nor how this process affects the gas of the central (Anglés-Alcázar et al., 2017; #2, 3). The 2020s will see advances here, especially at the low-mass end, as deeper and redder surveys will probe lower down the luminosity function in the outer reaches of the Local Volume—and around other galaxies at earlier epochs, placing the Milky Way's satellites in a broader context. But **the immediate environment of satellite galaxies is the CGM of a more massive galaxy**. For instance, star-forming centrals tend to have star-forming satellites, and likewise for quenched galaxies (Weinmann et al., 2006). While this "galactic conformity" is often attributed to assembly bias (Wechsler & Tinker, 2018), the *mechanism* is likely regulated by the CGM (#4). The Local Group contains the faintest known satellites, but CGM physics is necessary for interpreting the unique cosmological and baryonic effects present in these galaxies (Zolotov et al., 2012; #6, 7). It is thus critical to understand the structure and dynamics of this ambient gas.

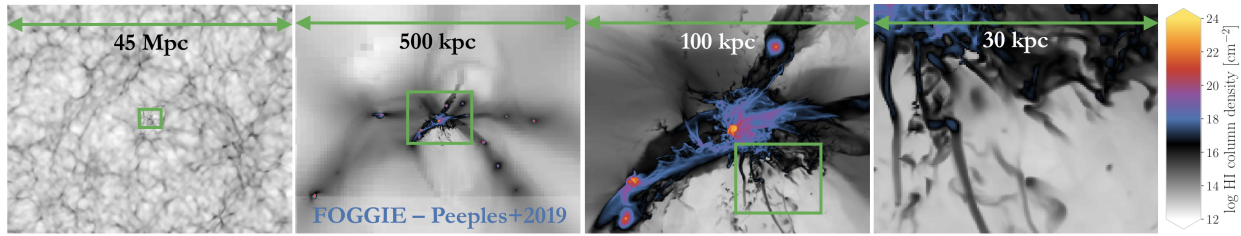

Figure 2: Neutral hydrogen modeled from the cosmic web to the small-scale CGM (Peeples et al., 2019); in the 2020s, these kinds of simulations need to reach smaller scales in larger volumes.

**6. The Dark Matter Halo Connection:** Wide-field ground- and space-based surveys of the 2020s will yield galaxy luminosity functions at both faint and bright magnitudes and to higher redshifts (#7) across a wide range of environments, further revealing how galaxies of different masses and star formation rates (SFRs) occupy halos throughout the cosmic web—vital both for understanding how galaxies evolve and for using them as cosmological probes. The luminosity function encapsulates the co-evolution between galaxies and their dark matter halos; differences in the number densities halos and galaxies speaks to star-formation–suppressing feedback mechanisms (cf. #5). Feedback is also invoked to explain the mismatch between galaxy SFRs and halos' baryonic accretion rates (Behroozi et al., 2013; Popping et al., 2015). **The relation between the halo accretion rate to SFR is modulated by the hundreds of kpc of CGM gas from the halo "edge" and the ISM**; whether the baryonic-to-dark matter accretion rate into the halo is at the cosmic fraction depends on, e.g., the CGM's pressure profile. While the luminosity function tells us the "quantity" of feedback required to reconcile galaxy and dark matter number densities, we ultimately must understand galaxies' extensive *gaseous* halos to understand the details of this connection.

**7. Reionization:** A long-awaited question the next decade promises to address is how the early universe was reionized. As the photons responsible for reionization will be unobservable, we will require constraining galaxies' fluxes of ionizing photons, what fraction of these photons escape into the IGM, and how the escape fraction varies with galaxy properties (mass, SFR, metallicity). Currently, reasonable differences in SPS models can lead to significant differences in the ionizing flux (Stanway et al., 2016), but this situation should improve in the 2020s. Yet **the fraction of galaxies' ionizing flux the *inter*galactic medium sees during reionization is modulated by the physical and thermal structure of the *circum*galactic medium**. Alarmingly, simulations show that merely better resolving the CGM—with no changes whatsoever to "subgrid" physics— significantly increases the CGM covering fraction optically-thick to the Lyman limit (van de Voort et al., 2019; Peeples et al., 2019). Faint satellite galaxies can provide independent constraints on reionization (e.g., Tollerud & Peek, 2018), but interactions with the CGM scrambles this information (#5). Thus to fully track reionization requires understanding small-scale CGM physics.

**What we will need to learn about the circumgalactic medium in the 2020s and beyond**

We list here four specific advances in CGM studies that must occur in order to address the above topics. Advances A–C are primarily observational, while D is primarily computational.

**A. We must spatially resolve the CGM—at high spectral resolution:** Most CGM data are individual "pencil beam" absorption measurements lacking the wealth of detail in the underlying gas distributions, kinematics, and abundances. Line-of-sight position and velocity are inherently degenerate: gas that is not co-spatial can fall at the same projected velocity, and gas accelerations can cause small spatial scales to map to large velocity ranges (Peeples et al., 2019). This degeneracy hampers ionization modeling and hence, e.g., metallicity estimates, and is exacerbated by low spectral resolution (Tripp & Bowen, 2005). CGM "maps" (e.g., COS-Halos; Tumlinson et al., 2013) constructed by aggregating single sightlines are also inherently incomplete. **The future lies in moving toward data that combine spatial sampling and high spectral resolution.** Detailed analyses with COS show multiphase kinematic structures in CGM gas that will require higher spectral resolution ($R \gtrsim 50,000$ or $\lesssim 6 \, \text{km s}^{-1}$) to properly disentangle (Werk et al., 2016). CGM gas flows can be resolved spatially for *individual* halos via a higher spatial density of background sources (Chen et al., 2014; Bowen et al., 2016; The LUVOIR Team, 2018) or via spatially-resolved sources (e.g., lensed galaxies; Rubin et al., 2018; Lopez et al., 2018). Emission maps would also constrain the CGM both spatially and kinematically (Corlies et al., 2019). Instruments mapping extended absorption and/or emission in the visible and NIR already exist (Keck/KCWI, VLT/MUSE, SDSS/MaNGA); extending these capabilities to the observed-frame UV and X-ray with wide fields and subarcsec resolution will bring these benefits to the local Universe.

**B. We must characterize the CGM of many galaxies of many sizes and types at many epochs:** Observations over the last decade have shown that "the CGM" is not monolithic; its composition, extent, and kinematic structure differ with galaxy properties and observed epoch. Yet even the largest current datasets probe only a few dozen galaxies, preventing in depth, multi-parameter characterization; **we must reach far fainter limits than accessible with current spectrographs**. For the last $10+$ billion years of cosmic time ($z \lesssim 2$), this advance further requires accessing ultraviolet and X-ray wavelengths (see, e.g., Figure 6 of Tumlinson, Peeples, & Werk, 2017).

**C. We must map the CGM with respect to spatially-resolved *galaxy* properties:** Mapping the full baryon cycle will require spatially resolving gas flows both in the CGM and *within* galaxies (Bordoloi et al., 2011; Borthakur et al., 2016; Martin et al., 2019). High spatial resolution maps of galaxies' gaseous and stellar kinematics and metallicities, in conjunction with circumgalactic maps, will constrain how galaxies gain angular momentum and redistribute metals. This will require **deep spatially-resolved spectroscopy at visible+NIR wavelengths** to map the stars and ionized gas and in the radio to map galaxies' neutral and molecular gas.

**D. We must model the CGM on small *and* cosmological scales:** The 2010s saw extraordinary advances in the ISM resolution in galaxy simulations that, coupled with more realistic star-formation driven feedback, led to more realistic simulated galaxies (e.g., Hopkins et al., 2014, 2018). **The 2020s will require a similar revolution in modeling the small-scale structure of the CGM in cosmological volumes** (van de Voort et al., 2019; Peeples et al., 2019; Corlies et al., 2019; Hummels et al., 2018; Suresh et al., 2019; Figure 2), necessitating not only codes that can effectively scale to the exascale, but also next-generation supercomputing facilities with high memory-per-core nodes. Extremely high-resolution idealized simulations yielding insights into small-scale gastrophysics (Armillotta et al., 2017; Fielding et al., 2018) will need be connected back to the cosmic web and self-consistent galaxy growth in order to place them in an observational context.

*Cover pages image adapted from Figure 1 of Tumlinson, Peeples, & Werk (2017), with gratitude to STScI's Ann Feild.*